\documentclass[prb,twocolumn,showpacs,preprintnumbers,amsmath,amssymb,superscriptaddress,floatfix]{revtex4-2}

\usepackage{amsmath}    
\usepackage{amssymb,environ}
\usepackage{graphicx}   
\usepackage{verbatim}   
\usepackage{color}      
\usepackage{hyperref}   
\usepackage{breakurl}
\usepackage[normalem]{ulem}
\usepackage{natbib}
\usepackage{enumitem}
\usepackage{multirow}

\usepackage{epsfig}
\usepackage{textcomp}
\usepackage{wasysym}
\usepackage{array}
\usepackage{color}

\begin{document}

\newcommand{\ev}[0]{\mathbf{e}}
\newcommand{\cv}[0]{\mathbf{c}}
\newcommand{\fv}[0]{\mathbf{f}}
\newcommand{\Rv}[0]{\mathbf{R}}
\newcommand{\Tr}[0]{\mathrm{Tr}}
\newcommand{\ud}[0]{\uparrow\downarrow}
\newcommand{\du}[0]{\downarrow\uparrow}
\newcommand{\Uv}[0]{\mathbf{U}}
\newcommand{\Iv}[0]{\mathbf{I}}
\newcommand{\Hv}[0]{\mathbf{H}}
\newcommand{\kv}[0]{\mathbf{k}}
\newcommand{\qv}[0]{\mathbf{q}}

\setlength{\jot}{2mm}

\newcommand{\jav}[1]{{\color{red}#1}}
\newcommand{\ds}[1]{{\color{blue}#1}}

\title{Spin susceptibility in a two-dimensional altermagnet with Hatsugai-Kohmoto interaction}
\title{Dynamical spin susceptibility of  $d$-wave Hatsugai-Kohmoto altermagnet}
\title{Interaction enhanced altermagnet in the Hatsugai-Kohmoto model}

\author{\'Ad\'am B\'acsi}
\email{bacsi.adam@sze.hu}
\affiliation{Department of Mathematics and Physics, Sz\'echenyi Istv\'an University, 9026 Gy\H or, Hungary}
\affiliation{MTA-BME Lendület "Momentum" Open Quantum Systems Research Group, Institute of Physics, Budapest University of Technology and Economics, Műegyetem rkp. 3., H-1111, Budapest, Hungary}

\author{Bal\'azs D\'ora}
\email{dora.balazs@ttk.bme.hu}
\affiliation{Department of Theoretical Physics, Institute of Physics, Budapest University of Technology and Economics, M\H uegyetem rkp. 3., H-1111 Budapest, Hungary}
\affiliation{MTA-BME Lendület "Momentum" Open Quantum Systems Research Group, Institute of Physics, Budapest University of Technology and Economics, Műegyetem rkp. 3., H-1111, Budapest, Hungary}

\begin{abstract}
We investigate the interplay between altermagnetic band structures and electronic correlations by 
focusing on the $d_{x^2-y^2}$ altermagnetic generalization of the Hatsugai-Kohmoto model.  
We find that with increasing interaction, a many-body Lifshitz transition takes place when doubly 
occupied regions disappear from the Fermi surface and almost all  momentum states become fully spin polarized.
This is termed interaction enhanced altermagnet.
We find that the dynamical susceptibility, which possesses only transverse non-zero components for small wavevectors, 
develops a gap proportional to the interaction strength, and displays a sharp peak at a frequency increasing with the interaction. 
Above the Lifshitz transition,
this peak moves to the lower gap edge and becomes log-divergent.
The signal intensity increases with the interaction up until the Lifshitz transition and saturates afterwards.
The static susceptibility remains unaffected by the correlations and altermagnetism
reduces the static transverse response.
\end{abstract}

\maketitle

\section{Introduction}

Altermagnets are a recently discovered class of magnetic materials~\cite{PRXSmejkal2022,smejkal2022,hayami2019,hayami2020,seebackalter2023,peng2024,purnendu2024, Krempasky2024, Gomonay2024,bai2024,Yang2025,Cheong2025,scalingalter2025,nenrstalter2025,song2025} characterized 
by simultaneous breaking of certain crystal symmetries and time reversal symmetry. The net magnetization of these materials is zero just like in antiferromagnets but they have a peculiar, spin-dependent band structure which is more 
typical to ferromagnets. The signatures of altermagnetism  are observable in numerous physical quantities such as the optical conductivity~\cite{peng2024} or angle-resolved spectroscopy measurements~\cite{gondolf2025,Yang2025} revealing 
significant spin-dependence of the band structure. Further experimental evidence comes from magneto-optical Kerr responses~\cite{iguchi2025} and anisotropic magnetoresistance~\cite{betancourt2024}, as well as from magnon transport measurements~\cite{hoyer2025}, all indicating robust spin polarization without macroscopic magnetization.
Due to their spin-dependent fermionic excitations, they are promising candidates for  information technology and spintronical applications.

From a theoretical perspective, altermagnetism arises from the interplay between collinear antiferromagnetic order and crystal symmetries, leading to alternating spin polarization in momentum space~\cite{PRXSmejkal2022,leeb2024}. This effect leads to a description in terms of noninteracting fermionic excitations with spin-dependent energy spectra. 
While this single-particle picture captures the essential band features, the role of electronic correlations on top of altermagnetism remains largely 
unexplored\cite{purnendu2024,kaushal}. Strong interactions often yield to a variety of peculiar phenomena, ranging from
Mott insulating behaviour and phase transitions to  non-Fermi liquid and pseudogap physics\cite{sachdev,giamarchi,mahan,Giuliani2005}.
A natural way to incorporate interactions into a tractable theoretical framework is provided by the Hatsugai-Kohmoto (HK) 
interaction~\cite{Phillips2020,hatsugai1992,lidsky,zhao2022,Zhao2023,Guerci2025,Ma2025,Zaanen2020}, 
which combines a transparent physical picture with  analytical solvability and which is at the same time a prototypical example of
 non-Fermi liquids~\cite{Zhao_2025}. In addition, the HK model\cite{Mai2025} can be continuously deformed to the Hubbard model by 
systematically introducing momentum mixing, going from qualitative to quantitative agreement between these models. 

Here we focus on the effect of strong correlations  in an altermagnet. 
We employ a direct evaluation of the Kubo formula~\cite{Mahan2000,Kubo1957} by explicitly incorporating many-body occupation probabilities,
which represents a natural framework to study the response functions of the HK model. 
This is in contrast to the conventional treatment based on single-particle Green's function and single-particle occupation probabilities. Therefore, 
our approach provides a more intuitive and suitable basis to analyze the interplay between altermagnetism and the HK interaction.   

We identify rich physics, including a many-body Lifshitz transition due to correlations, when momentum space doublons are excluded from the Fermi surface and
almost all momentum states become fully spin polarized. This results in interaction enhanced altermagnetism.
This is also reflected in the dynamical spin response. It displays an interaction induced gap and a sharp peak, which becomes divergent above the  Lifshitz
transition. In this regime, the signal intensity remains constant in the interaction enhanced altermagnetic phase due to the frozen and spin polarized Fermi surface without doublons, only its lower gap edge moves to higher frequencies with increasing interaction.
 In spite of the strong interaction dependence of the dynamical susceptibility, the static, Pauli response remains independent from correlations and is
only influenced by the altermagnetic band structure.

\section{Altermagnet with Hatsugai-Kohmoto interaction}

Altermagnets are known for their spin-dependent energy spectra as a consequence of lifted Kramers spin-degeneracy~\cite{Krempasky2024}. In this paper, we study the interplay between the spin-dependence of a two-dimensional altermagnet
and the Hatsugai-Kohmoto (HK) interaction describing a local interaction in momentum space. The Hamiltonian of the system reads
\begin{gather}
H=\sum_{\kv s}\varepsilon_s(\kv)c_{\kv s}^+ c_{\kv s} + \sum_{\kv}Un_{\kv\uparrow}n_{\kv\downarrow},
\label{eq:ham}
\end{gather}
where the spin-dependent energy spectrum
\begin{gather}
\varepsilon_\uparrow(\kv) = \frac{ k_x^2}{2m} ~ \alpha  + \frac{ k_y^2}{2m} ~ \frac{1}{\alpha}, \hspace*{3mm}
\varepsilon_\downarrow(\kv) = \frac{ k_x^2}{2m} ~ \frac{1}{\alpha}  + \frac{ k_y^2}{2m} ~ \alpha
\label{eq:spectra}
\end{gather}
describe the $d_{x^2-y^2}$ altermagnetism with $\alpha$ measuring the asymmetry between band structures of the opposite 
spin directions ($\alpha=1$ corresponds to the symmetric case). We also note that our results also apply to altermagnets with $d_{xy}$ symmetry. 
The energy spectrum \eqref{eq:spectra} describes the effective low-energy
model expanded to quadratic order in momenta around the spin-degeneracy points (Brillouin zone center) of systems obeying combined 
time-reversal and four-fold rotation symmetry, a typical feature in altermagnets \cite{smejkal2022,Krempasky2024,peng2024}.

In the second term of Eq. \eqref{eq:ham}, $U$ measures the strength of the HK interaction \cite{hatsugai1992,Phillips2020} and $n_{\kv s} = c_{\kv s}^+c_{\kv s}$ with $c_{\kv s}$ the 
annihilation operator of electrons and we use $\hbar = 1$.

Note that the system breaks both the time reversal $T$ and the rotation symmetry but preserves $C_4 T$, the combination of a fourfold rotation and time-reversal \cite{peng2024}.

We comment on the physical relevance of our model in Eq. \eqref{eq:ham}, which is
 deceivingly simple and contains a contact interaction in momentum space, which translates to a long range interaction in real space\cite{Phillips2020}. While this
is hardly realized in physical systems, the ensuing physics of our model is expected to describe qualitatively the behaviour of strongly correlated altermagnets. In addition to the altermagnetic
band structure, the interaction term penalizes double occupancies, which is the general behaviour of Hubbard type models. Moreover, the HK model\cite{Mai2025} was shown to be continuously connected
 to the Hubbard model by  introducing momentum mixing terms in the interaction, resulting in quantitative agreement between these models. Consequently, 
the HK model provides at least a qualitative description of Hubbard type physics.

As to possible other phases, the low energy spectrum in Eq. \eqref{eq:spectra} applies to low fillings on tight-binding models\cite{peng2024}, away from Fermi surface nesting conditions. 
Therefore,
no additional phases are expected for repulsive interactions\cite{leblanc}, except for the possible Mottness. Close to commensurate fillings, additional phases (e.g. antiferromagnetic) could be possible, which
is beyond the scope of our model and investigation.

The great advantage of the HK model is that the Hamiltonian decouples to different wavenumber sectors characterized by four many-body basis states of $|\kv,0\rangle$, $|\kv,\uparrow\rangle$, $|\kv,\downarrow\rangle$ and $|\kv,\ud\rangle$. 
These describe states with no electron, only spin up occupancy, only spin down occupancy or double occupancy in the mode $\kv$, respectively.
The density matrix at thermal equilibrium is diagonal in this basis with the entries of the following many-body occupation probabilities:
\begin{eqnarray}
P(\kv,0) & = & \langle (1-n_{\kv\uparrow})(1-n_{\kv\downarrow})\rangle = \frac{1}{Z_\kv}, \nonumber \\
P(\kv,s)  & = & \langle n_{\kv s}(1-n_{\kv\bar{s}})\rangle = \frac{e^{-\beta\xi_s(\kv)}}{Z_\kv}, \nonumber \\
P(\kv,\ud) & = & \langle n_{\kv\uparrow}n_{\kv\downarrow}\rangle = \frac{e^{-\beta(\xi_\uparrow(\kv) + \xi_\downarrow(\kv) + U)}}{Z_\kv}
\label{eq:Ps}
\end{eqnarray}
with $\bar{s}=-s$ and
\begin{gather}
Z_\kv = 1 + e^{-\beta\xi_\uparrow(\kv)} + e^{-\beta\xi_\downarrow(\kv)} + e^{-\beta(\xi_\uparrow(\kv) + \xi_\downarrow(\kv) + U)}
\end{gather}
the partition function corresponding to the momentum channel $\kv$, $\xi_s(\kv)=\varepsilon_s(\kv)-\mu$ and $\beta$ is the inverse temperature.
The probabilities $P(\kv,\sqcup)$ are the many-body counterparts of the products of occupation numbers in a typical non-interacting situation. In fact, when $U=0$, these occupation probabilities boil down to simple products with the Fermi function. 
For example, $P(\kv,\uparrow) = f(\xi_\uparrow(\kv)) (1-f(\xi_\downarrow(\kv)))$ for $U=0$.

\begin{figure}[h]
\centering
\includegraphics[width=8cm]{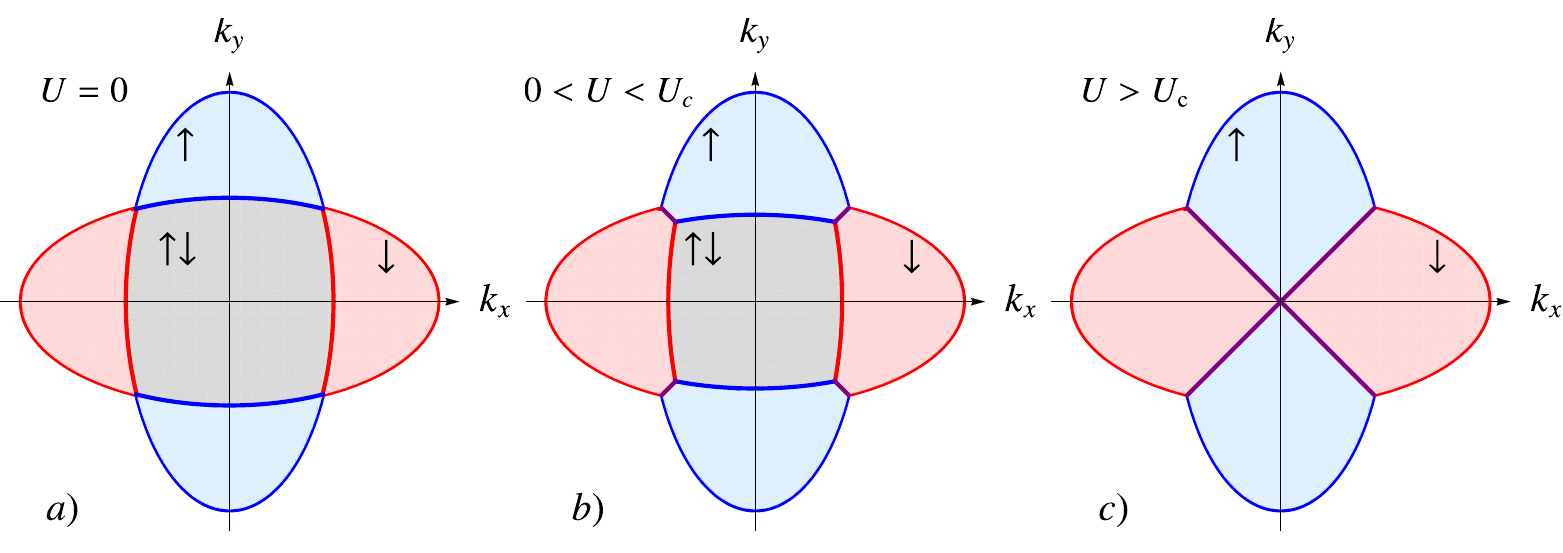}
\caption{Schematic picture of the ground state occupation of many-body states as a function of momentum in a) the non-interacting 
($U=0$),  b) moderately interacting  ($0<U<U_{c}$)
and c) strongly interacting ($U>U_{c}$) cases. In the blue region, the up spin state is occupied while the down spin state is empty.
In the red region, only the down spin state is occupied. The grey area indicates double occupancy. Panel c) can be termed "interaction enhanced altermagnet" due to the lack of double occupancies and fully polarized momentum states.}
\label{fig:kspace}
\end{figure}

Let us now examine the ground state of the system assuming that the electronic spectrum is filled up to the chemical potential $\mu$. 
In the ground state, the many-body occupation probabilities become either one, for the lowest energy many-body state, or zero, for the other states. 
In the non-interacting case, $U=0$, Fig. \ref{fig:kspace} a) shows the $\kv$ dependence of the spin-configuration realized in the ground state. The Fermi 
surface consists of two, intersecting ellipses corresponding to the two spin orientations. The peculiarity of the altermagnetic electron spectrum is 
that the ground state includes singly occupied regions with well-defined spin polarization, see the blue and red regions in Fig. \ref{fig:kspace} a). 

In the presence of finite interaction strength $U$, the double occupancy is suppressed at fixed chemical potential, 
see Fig. \ref{fig:kspace} b). The size of singly-occupied, spin-polarized regions grows, 
therefore, altermagnetic features are expected to be enhanced by increasing the interaction $U$. This can be called an \emph{interaction enhanced altermagnet}. We note that this is not an 
emergent structure but rather follows from the interplay of the original altermagnetic band dispersion and interaction.

At fixed number of electrons, a $U$-dependent chemical potential emerges as
\begin{gather}
\mu(U)=\left\{\begin{array}{cc}   
\left(1-\frac{U}{U_c}\right)\mu_0 +U & \mbox{for $U<U_{c}$} \\ U_{c} & \mbox{for }U>U_c \end{array}\right.,
\label{eq:muU}
\end{gather}
where $\mu_0= \frac{\pi n_e}{m}$ is the chemical potential at $U=0$ corresponding to the electron density $n_e$,
$U_{c} =\mu_0\left(1-B(\alpha) \right)^{-1} $ with $B(\alpha) = \frac{2}{\pi}\arctan\left(\mathrm{min}(\alpha,\alpha^{-1})\right)$.
Here, $U_c$ denotes the critical
interaction strength above which no double occupancy is present for a given number of 
particles in the ground state and each momentum state is spin polarized except 
for the momenta along the degeneracy curves at the border of the spin-up and spin-down regions which however constitute a zero-measure set in the momentum space,
 therefore, the system is almost everywhere spin-polarized. This we will refer to as fully spin-polarized state and it is shown in Fig. \ref{fig:kspace} c).  
The exclusion of double occupancies in the Fermi surface indicates a many-body Lifshitz transition\cite{lifshitz,volovik} at $U_{c}$.
In the $U>U_{c}$ region, there still exist gapless excitations but the system exhibits Mott insulator features in the sense that the shape of the Fermi 
surface does not change with further increase of $U$ \cite{Phillips2020} and optical excitations do not probe the gapless modes. Hence, this regime can be regarded as an effective Mott insulating altermagnet. 
A real Mott insulator without additional gapless modes is expected to emerge at specific fillings beyond the low-energy model, 
when the interaction strength becomes comparable to the bandwidth.

The Lifshitz transition is not a typical weak-coupling instability because its critical interaction range is $\mu_0<U_c<2\mu_0$ depending on the value of $\alpha$.
This is further corroborated by evaluating the density of particles in doublonic configuration in the ground state from 
\begin{gather}
n_{\ud}=\frac{2}{A} \sum_k P(\kv,\ud)=2n_e\left(1-\frac{U}{U_{c}}\right)B(\alpha),
\end{gather}
when $U<U_{c}$ and zero for $U>U_{c}$, $A$ is the area of the system. For $U=0$ and $\alpha=1$, the ground state consists of momentum space
doublons and $n_{\ud}=n_e$.

The analysis of the ground state shows that the HK interaction reorganizes the Fermi surface and in turn, the structure of occupation probabilities 
which is dominated by many-body features. 
In the followings, we study how these correlation effects influence the linear response of the system.

\section{Spin susceptibility tensor}

The unique signature of altermagnets is their peculiar spin dependent band structure in Fig. \ref{fig:kspace}. The best way to directly probe and reveal this is to study spin
dependent quantities, such as the spin response. To this end, we analyze the full $3\times 3$ spin susceptibility tensor which is defined by the Kubo formula as
\begin{gather}
\chi_{nm}(\Rv,t)=i\mu_B^2\left\langle \left[\hat{S}^{n}(\Rv,t),\hat{S}^{m}(0,0)\right]\right\rangle
\label{eq:kubo}
\end{gather}
for positive times and 0 for negative $t$.
The indices $n$ and $m$ stand for the spin components $x,y$ or $z$, $\mu_B$ is the Bohr magneton and $\Rv$ is the two-dimensional space-variable. The spin operators are expressed with the fermionic operators as
\begin{gather}
\hat{S}^{n}(\Rv,t) = \frac{1}{N}\sum_{\kv\kv',ss'}e^{i(\kv-\kv')\Rv}c_{\kv s}^+(t) \sigma_{ss'}^n c_{\kv' s'}(t)
\label{eq:spinoperator}
\end{gather}
with $\sigma^n$ denoting the Pauli matrices and $N$ the number of unit cells. 
After some algebraic steps, detailed in Appendix, we obtain
\begin{widetext}
\begin{gather}
\chi_{nm}(\qv,\omega) = \lim_{\delta\rightarrow 0^+}\int_{0}^\infty e^{i\omega t-\delta t} \chi(\qv,t)\,\mathrm{d}t = 
-\frac{\mu_B^2}{N}\delta_{\qv,0}\sum_{\kv,s}\sigma_{s\bar{s}}^n \sigma_{\bar{s}s}^m \frac{P(\kv,s) - P(\kv,\bar{s})}{ \omega + i\delta + \varepsilon_{s}(\kv) - \varepsilon_{\bar{s}}(\kv)}
-\nonumber \\ -
\frac{\mu_B^2}{N}\left(1-\delta_{\qv,0}\right)\sum_{\kv,ss'} \sigma_{ss'}^n \sigma_{s's}^m \left[\frac{P(\kv + \qv,s)P(\kv,0) - P(\kv + \qv,0)P(\kv,s')}{\omega + i\delta + \varepsilon_{s}(\kv + \qv) - \varepsilon_{s'}(\kv)} + \right. \nonumber \\ + 
\frac{P(\kv + \qv,\ud)P(\kv,0) - P(\kv + \qv,\bar{s}) P(\kv,s')}{\omega + i\delta + \varepsilon_{s}(\kv + \qv) - \varepsilon_{s'}(\kv)+U} +
\frac{P(\kv + \qv,s)P(\kv,\bar{s}') - P(\kv + \qv,0)P(\kv,\ud)}{\omega + i\delta + \varepsilon_{s}(\kv + \qv) - \varepsilon_{s'}(\kv) - U} +
\nonumber \\ + \left.
\frac{P(\kv + \qv,\ud)P(\kv,\bar{s}') - P(\kv + \qv,\bar{s})P(\kv,\ud)}{\omega + i\delta + \varepsilon_{s}(\kv + \qv) - \varepsilon_{s'}(\kv)}\right],
\label{eq:chinm}
\end{gather}
\end{widetext}
which is the most general form of the spin susceptibility tensor for a system with $\kv$-diagonal many body occupation probabilities.

The first term in the susceptibility corresponds to the exact zero wavenumber, $\qv=0$. However, this contribution is physically not 
relevant because all excitations, including optical excitations, possess a non-zero wavenumber. Therefore, this term will be neglected in the followings similarly to Ref. \cite{Zhao2023}.
We note that, due to the properties of the Pauli matrices, the tensor elements $\chi_{xz}$, $\chi_{zx}$, $\chi_{yz}$ and $\chi_{zy}$ vanish. 
The remaining tensor elements will be studied in specific cases.

\section{Dynamical spin susceptibility}

We consider the homogeneous limit, $\qv\rightarrow 0$, while keeping $\omega$ finite. This is relevant experimentally in condensed matter\cite{blundell} 
for neutron\cite{maier,faure} as well as X-ray
scattering, electron spin resonance studies\cite{Slichterbook,Sun2025,povarov}
 and for cold atoms using  spin-dependent Bragg spectroscopy\cite{stenger}. 

 We focus on  zero temperature such that the probabilities $P(\kv,\sqcup)$ take values of 1 inside certain regions of the momentum space and 0 outside. This significantly simplifies Eq. \eqref{eq:chinm}, as the products $P(\kv,\sqcup)P(\kv,\sqcup')$ vanish whenever $\sqcup$ and $\sqcup'$ are different as these are mutually exclusive events or probabilities. 
As a consequence, the tensor element $\chi_{zz}(\omega)$ is identically zero. The remaining components are expressed as
\begin{gather}
\chi_{xx}(\omega) = \chi_{yy}(\omega) = \Gamma_{\ud}(\omega) + \Gamma_{\du}(\omega), \nonumber \\
\chi_{xy}(\omega) = -\chi_{yx}(\omega) = i\left(\Gamma_{\ud}(\omega) - \Gamma_{\du}(\omega)\right),
\end{gather}
where
\begin{gather}
\Gamma_{\ud}(\omega) =
 \frac{\mu_B^2}{N}\lim_{\delta\rightarrow 0^+}\sum_{\kv} \left[ 
\frac{P(\kv,\uparrow)^2 }{\omega + i\delta + \varepsilon_{\downarrow}(\kv) - \varepsilon_{\uparrow}(\kv) + U} - \right. \nonumber \\ - \left.
\frac{P(\kv,\downarrow)^2 }{\omega + i\delta + \varepsilon_{\downarrow}(\kv) - \varepsilon_{\uparrow}(\kv) - U}\right],
\label{eq:zero}
\end{gather}
and $\Gamma_{\du}(\omega)= (\Gamma_{\ud}(-\omega))^*$ are the spin-flip response functions. The non-vanishing terms in $\Gamma_{\ud}(\omega)$ can be associated with specific spin-flip processes. The first term with $P(\kv,\uparrow)^2$ corresponds to the situation where two spin-up particles occupy states at $\kv$ and infinitesimally close to $\kv$ within the blue region of Fig. \ref{fig:kspace} b). In this region, the spin-down state is unoccupied, allowing the spin-flip process resulting in double occupancy at $\kv$. Similarly, the second term $P(\kv,\downarrow)^2$ corresponds to a spin-flip process involving spin-down particles in the red region of Fig. \ref{fig:kspace} b).

After taking the limit $\delta\rightarrow 0^+$, two kinds of terms appear in the response function, $\Gamma_{\ud}(\omega) = \Gamma_{\ud}'(\omega) + i\Gamma_{\ud}''(\omega)$. Here, $\Gamma'$ includes the terms with the principal value functions and $\Gamma''$ includes the Dirac-delta terms. Note that $\Gamma'$ and $\Gamma''$ are related to each other by Kramers-Kronig relations. Henceforth, we focus on $\Gamma''$ only.

The frequency dependence is analytically obtained as
\begin{widetext}
\begin{gather}
\Gamma_{\ud}''(\omega) = \frac{\mu_B^2 g(\varepsilon_F)}{\alpha - \alpha^{-1}}
\left\{\begin{array}{ll}
\mathrm{artanh}\sqrt{\frac{(\alpha^2-1)\mu - (\omega - U)}{(\alpha^2-1)\mu + \alpha^2 (\omega - U)}}
- \mathrm{artanh}\sqrt{\frac{(\alpha^2-1)(\mu-U) - \alpha^2(\omega - U)}{(\alpha^2-1)(\mu-U) + (\omega - U)}} & \mbox{for $U< \omega <\omega_m$} \\
\mathrm{artanh}\sqrt{\frac{(\alpha^2-1)\mu - (\omega - U)}{(\alpha^2-1)\mu + \alpha^2 (\omega - U)}} & \mbox{for $\omega_m<\omega <\alpha^2\omega_m$} 
\end{array}\right.
\label{eq:Som}
\end{gather}

\end{widetext}
and 0 otherwise for $\omega>0$ and $\Gamma_{\ud}''(-\omega) = - \Gamma_{\ud}''(\omega)$ for the negative frequencies. In Eq. \eqref{eq:Som}, 
\begin{gather}
\omega_m=U+\left(1-\frac{1}{\alpha^{2}}\right)(\mu-U)
\label{eq:omm}
\end{gather} 
denotes the location of the peak with the most intense spin-flip response, 
 $g(\varepsilon_F)=\frac{A_c m}{\pi }$
is the total density of states of the non-interacting electrons with $A_c=A/N$ the area of the unit cell and includes both spin directions. It is also energy and $\alpha$ independent.
Eqs. \eqref{eq:Som} and \eqref{eq:omm} are valid for $\alpha>1$, and for $\alpha<1$, $\alpha$ should be replaced by $\alpha^{-1}$.

\begin{figure}[h]
\centering
\includegraphics[width=8cm]{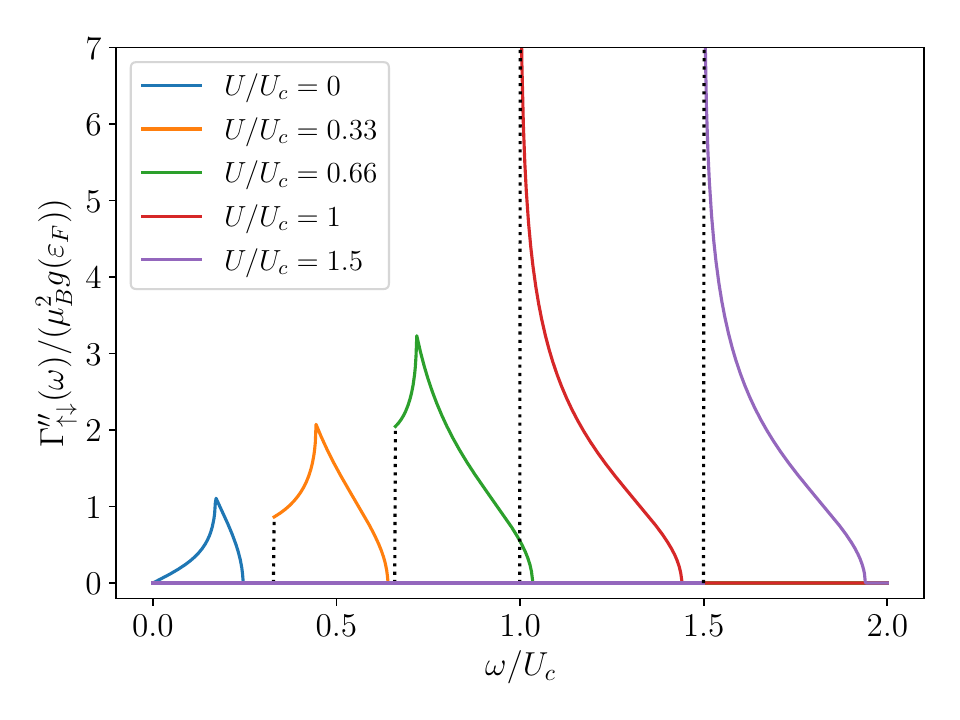}
\caption{Frequency dependence of the dynamical spin susceptibility is shown at $\alpha = 1.2$ for various values of $U$ and $\mu_0=(1-B(1.2))U_{c}$. }
\label{fig:gammaud}
\end{figure}

Due to the oddness of $\Gamma_{\ud}''(\omega)$, the tensor elements $\chi_{xy}''(\omega)$ and $\chi_{yx}''(\omega)=0$ vanish while 
$\chi_{xx}''(\omega) = \chi_{yy}''(\omega)=2\Gamma_{\ud}''(\omega)$. The frequency dependence of $\Gamma_{\ud}''$ is plotted in Fig. \ref{fig:gammaud}.

The spin susceptibility develops a gap of $U$ at finite interaction strength. This is the energy cost of forming a $\ud$ pair from two spin-up particles located infinitesimally close to each other through the spin-flip process. 
Fig. \ref{fig:gammaud} shows that increasing interaction strength enhances the signal intensity of the dynamical spin susceptibility which also originates from  the enlarged spin-polarized regions in the momentum space from Fig. \ref{fig:kspace}. The maximum response at $\omega_m$ is found as
\begin{gather}
\Gamma_{\ud,max}'' = \frac{\mu_B^2 g(\varepsilon_F)}{\alpha - \alpha^{-1}}\mathrm{artanh}\left( \sqrt{\frac{\mu(\alpha^2-1) + U}{\alpha^2(2\mu - U)}}    \right)\,.
\label{eq:Gudmax}
\end{gather}
As the interaction strength approaches $U_{c}$, the maximum location shifts to the lower gap edge $\omega_m = U$ and the maximum 
value, Eq. \eqref{eq:Gudmax} diverges logarithmically.

Beyond the Lifshitz transition, $U>U_{c}$, the frequency dependence has the same structure as at $U_{c}$ with a log divergent peak at the lower gap edge as
$\sim-\ln(\omega-U)$. This arises from a saddle point in momentum space from the denominator of Eq. \eqref{eq:zero} 
since $\varepsilon_\uparrow(\kv)-\varepsilon_\downarrow(\kv)\sim k_x^2-k_y^2$.

The dynamical susceptibility above the Lifshitz transition is described by the second row of Eq. \eqref{eq:Som} with $\mu = U_{c}$  and is 
located between $U<\omega<U + (\alpha^2-1)U_{c}$. Hence, the shape of the response function remains essentially unchanged. 
Further increasing the interaction strength 
 only shifts the response function toward higher frequencies, reflecting the growing separation between the lower and upper Hubbard bands. 
The shape of the Fermi surface does not change any more in Fig. \ref{fig:kspace} as doublons are already excluded from it.
Similar features are expected to occur in an altermagnetic Mott insulators beyond the transition point. There, the lower Hubbard band is 
completely filled\cite{Phillips2020} 
without doublons and is separated by a clean gap from the upper Hubbard band. Although our lower Hubbard band is only partially filled in this case, 
small momentum optical excitations only allow for vertical transitions and cannot reveal the partial filling of  the lower band. Therefore,
the $U>U_{c}$ small momentum spin response is analogous in this respect to that of an altermagnetic Mott insulator. 

Finally, we comment on the nature of gapped spin excitations in an otherwise gapless system. The linear shift of the resonance peak upon increasing interaction strength may be model dependent but the gap opening is expected on general ground 
due to the extra energy associated to double occupancies from the interaction.

\section{Static susceptibility}

In this section, we analyze the static susceptibility, $\omega=0$, in the homogeneous limit, $\qv\rightarrow 0$. In contrast to the dynamical susceptibility, the $zz$ component of the static response function remains finite at zero temperature. Interestingly, this component is independent of the interaction strength $U$ and the asymmetry parameter $\alpha$, and is given by
\begin{gather}
\chi_{zz}^{st} = 
\mu_B^2 g(\varepsilon_F),
\end{gather}
which is the conventional Pauli susceptibility\cite{abrikosov}.

\begin{figure}[h]
\centering
\includegraphics[width=8cm]{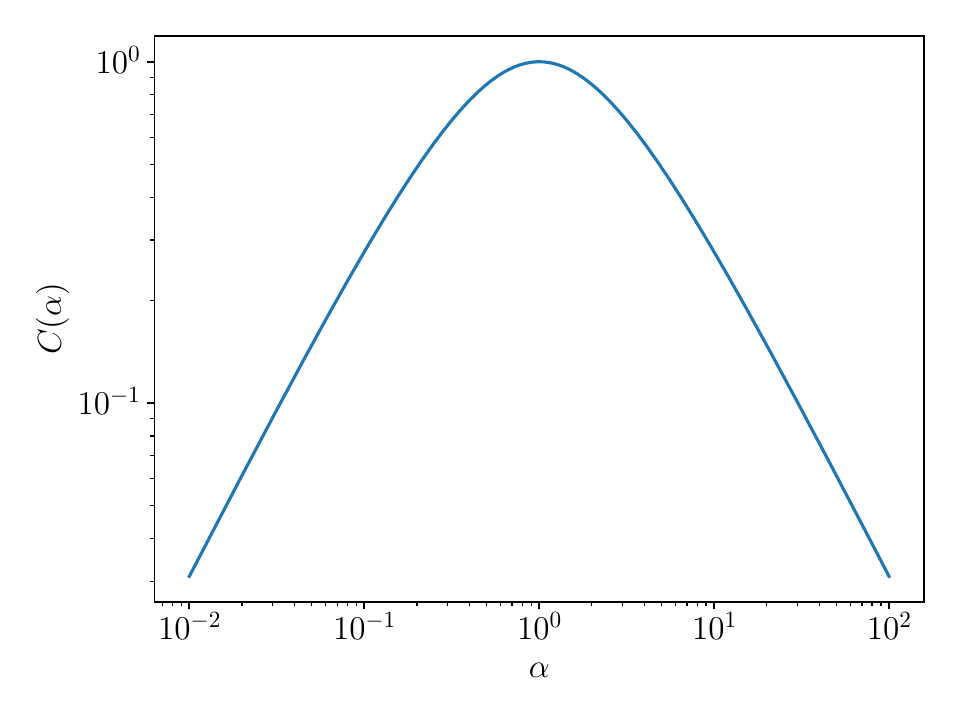}
\caption{$C(\alpha)$ is plotted for the $\alpha$ dependence of the transverse static spin susceptibility, which is symmetric for $\alpha\leftrightarrow 1/\alpha$ change from Eq. \eqref{calpha}.}
\label{fig:Cfun}
\end{figure}

The remaining, non-vanishing components of the susceptibility tensor are calculated as 
\begin{gather}
\chi_{xx}^{st}= \chi_{yy}^{st} = \chi_{zz}^{st}~ C(\alpha,U)
\end{gather}
with
\begin{gather}
C(\alpha,U) = \frac{4}{\pi} \int_{\pi/4}^{\pi/2} \frac{\mathrm{d}\varphi}{v(\varphi)}  \ln\left(\frac{\frac{U}{\mu} + \frac{2v(\varphi)}{\alpha + \frac{1}{\alpha} - v(\varphi)}}{\frac{U}{\mu} + \frac{2\left(1-\frac{U}{\mu}\right)v(\varphi)}{\alpha + \frac{1}{\alpha} + v(\varphi)}}   \right),
\end{gather}
where $v(\varphi) = \cos(2\varphi)\left(\frac{1}{\alpha}-\alpha \right)$. After some algebraic steps, it can be shown that the integral is independent of the interaction strength $U$. This independence 
is related to the fact that contributions to the static susceptibility dominantly come from intraband processes. Since the finite interaction only splits the band structure to an upper and lower Hubbard 
band \cite{Phillips2020} and does not alter its shape, the characteristics of intraband processes are expected to be preserved at finite interaction strength. The $U$-independent integral
\begin{gather}
C(\alpha) = \frac{4}{\pi} \int_{\pi/4}^{\pi/2} \frac{\mathrm{d}\varphi}{v(\varphi)}  \ln\left(\frac{\alpha + \frac{1}{\alpha} + v(\varphi)}{\alpha + \frac{1}{\alpha} - v(\varphi)}   \right)  
\label{calpha}
\end{gather}
is computed numerically and is symmetric under $\alpha\leftrightarrow 1/\alpha$. The results shown in Fig. \ref{fig:Cfun} indicate that the asymmetry parameter $\alpha$ suppresses the susceptibility as moving away from the symmetrical point $\alpha=1$.
In particular, $C(\alpha\rightarrow\infty)\rightarrow\pi/\alpha$.

\section{Discussion}

We have analyzed the interplay between altermagnetism and electronic correlations through the lens of the spin susceptibility. 
The electron-electron interaction has been modelled using the Hatsugai-Kohmoto model which enables analytically exact solutions. It serves as baseline model for non-Fermi liquids and Mottness and by introducing proper momentum mixing, can also be continuously deformed to the Hubbard model\cite{Mai2025}. 
The exact effects of the momentum-mixing terms on the spin response function need further investigations but
qualitatively similar results are expected for the altermagnetic Hubbard model at low fillings.
By studying a Hatsugai-Kohmoto altermagnet, we find that the non-interacting altermagnet undergoes a many-body Lifshitz transition with increasing
interaction when momentum space doublons are completely excluded from the Fermi surface and almost all  momentum states are fully spin polarized.
The ensuing state is an interaction enhanced altermagnet.
The spin susceptibility is evaluated using the many-body occupation probabilities. The dynamical spin susceptibility develops a gap proportional to the interaction
and initially displays a sharp though finite peak for larger frequencies.

Above the Lifshitz transition, this peak moves to the lower gap edge and becomes log-divergent due to a momentum space saddle point.
There, the Fermi surface remains intact and the shape of the dynamical spin susceptibility does not change further but acquires
an overall shift to larger frequencies.
Parallel to these, the signal intensity grows with interaction up until the transition and remains unchanged by further increasing the interaction.

The static, Pauli limit of the spin response is independent from the interaction strength though the transverse component gets altered by the 
altermagnetic band structure.
Our results indicate how strong electronic correlations can enhance altermagnetism and reveal its fingerprints on dynamical spin probes.

\section*{Data availability statement}
The data that support the findings of this article are openly available \cite{bacsi_2026_20176677}.

\begin{acknowledgments}
This work was supported by the National Research, Development and Innovation Office - NKFIH  Project Nos. K134437 and K142179,
by a grant of the Ministry of Research, Innovation and
 Digitization, CNCS/CCCDI-UEFISCDI, under projects number
PN-IV-P1-PCE-2023-0159 and PN-IV-P1-PCE-2023-0987.
This work was also supported by the HUN-REN Hungarian Research Network through the Supported Research Groups
Programme, HUN-REN-BME-BCE Quantum Technology Research Group (TKCS-2024/34).

\end{acknowledgments}

\bibliographystyle{apsrev}
\bibliography{hk}

\newpage

\appendix 

\section{On the derivation of Eq. \eqref{eq:chinm}}

Here we discuss the explicit evaluation  of the time dependent Kubo formula for the Hatsugai-Kohmoto model.
For Eq. \eqref{eq:spinoperator},
the time-dependence of fermionic annihilation operators is obtained analytically as
\begin{gather}
c_{\kv s}(t) = c_{\kv s} e^{-i\varepsilon_s(\kv)t}\underbrace{\left( 1 - \left(1 - e^{-iUt}\right) n_{\kv \bar{s}}\right)}_{a_{\kv\bar{s}}(t)},
\end{gather}
where we introduced the operator $a_{\kv \bar{s}}(t)$ involving the occupation number operator $n_{\kv\bar{s}}=c_{\kv\bar{s}}^+ c_{\kv\bar{s}}$.
Note that neither $a_{\kv \bar{s}}(t)$ nor $c_{\kv s}(t)$ commute with $c_{\kv\bar{s}}$. After substitution into the Kubo formula, Eq. \eqref{eq:kubo}, we have
\begin{gather}
\chi_{nm}(\Rv,t)=\frac{i\mu_B^2}{ N^2}\sum_{\kv\kv',ss'}\sum_{\kv_1\kv_1',s_1s_1'}e^{i(\kv-\kv')\Rv}e^{i\left(\varepsilon_s(\kv) - \varepsilon_{s'}(\kv')\right)t}
\times \nonumber \\
\sigma_{ss'}^n \sigma_{s_1s'_1}^m \left\langle \left[c_{\kv s}^+ a_{\kv \bar{s}}(t)^+a_{\kv' \bar{s}'}(t) c_{\kv's'},c_{\kv_1 s_1}^+ c_{\kv'_1 s'_1}\right]\right\rangle,
\end{gather}
where $\langle\rangle$ stands for the expectation value in thermal equilibrium. First we evaluate the commutators and the expectation value. The operators $a$ do not
change the set of particles in a state. Therefore, the creation operators $c^+$ must be paired with the annihilation operators $c$ in some order. Non-zero terms are
either proportional to $\delta_{\kv\kv'}\delta_{ss'}\delta_{\kv_1\kv_1'}\delta_{s_1s_1'}$ or to $\delta_{\kv\kv_1'}\delta_{ss_1'}\delta_{\kv_1\kv'}\delta_{s_1s'}$.
It can be shown that the $\delta_{\kv\kv'}\delta_{ss'}\delta_{\kv_1\kv_1'}\delta_{s_1s_1'}$ terms vanish. We substitute the remaining terms, which are proportional
to $\delta_{\kv\kv_1'}\delta_{ss_1'}\delta_{\kv_1\kv'}\delta_{s_1s'}$, into the susceptibility and perform spatial Fourier transformation.
\begin{gather} 
\chi_{nm}(\qv,t) = \sum_{\Rv}e^{-i\qv\Rv}\chi_{nm}(\Rv,t) = \nonumber \\ =
\frac{i\mu_B^2}{N}\sum_{\kv\kv',ss'}\delta_{\kv,\kv'+\qv}e^{i\left(\varepsilon_s(\kv) - \varepsilon_{s'}(\kv')\right)t}\sigma_{ss'}^n \sigma_{s's}^m C_{\kv\kv',ss'}(t),
\label{eq:ssusC}
\end{gather}
where
\begin{gather}  
C_{\kv\kv',ss'}(t) 
= \delta_{\kv\kv'}\delta_{s\bar{s}'}\Big\langle n_{\kv s}(1-n_{\kv\bar{s}}) - (1-n_{\kv s}) n_{\kv\bar{s}}  \Big\rangle + \nonumber \\
+ \left(1-\delta_{\kv\kv'}\right)
\Big\langle a^+_{\kv\bar{s}}(t) n_{\kv s}\Big\rangle \Big\langle a_{\kv' \bar{s'}}(t)(1-n_{\kv's'})\Big\rangle - \nonumber \\
- \left(1-\delta_{\kv\kv'}\right) \Big\langle a^+_{\kv\bar{s}}(t) (1-n_{\kv s})\Big\rangle \Big\langle a_{\kv' \bar{s'}}(t)n_{\kv's'}\Big\rangle \,.
\end{gather}
By using the probabilities $P(\kv,\sqcup)$ of Eq. \eqref{eq:Ps}, we obtain

\begin{gather}
C_{\kv\kv',ss'}(t) = \delta_{\kv\kv'}\delta_{s\bar{s}'}\left(P(\kv,s) - P(\kv,\bar{s})\right) + \nonumber \\
+ \left(1-\delta_{\kv\kv'}\right)
\Big[e^{iUt} \left( P(\kv,\ud)P(\kv',0) - P(\kv,\bar{s})P(\kv',s')\right) +  \nonumber \\
+ e^{-iUt}\left(P(\kv,s) P(\kv',\bar{s}') - P(\kv,0) P(\kv',\ud)\right) + \nonumber \\
+ P(\kv,s)P(\kv',0) + P(\kv,\ud)P(\kv',\bar{s}') - \nonumber \\ -   P(\kv,0) P(\kv',s') - P(\kv,\bar{s}) P(\kv',\ud)\Big]\,.
\end{gather}

We substitute into Eq. \eqref{eq:ssusC} and perform the temporal Fourier transformation leading to Eq. \eqref{eq:chinm}.

\end{document}